\documentclass[11pt,a4paper]{article}
\usepackage{proteus-paper}

\title{Proteus: Automated Adversarial Robustness Testing\\for Audio Deepfake Detectors}
\shorttitle{Proteus: Robustness Testing for Audio Deepfake Detectors}
\venue{International Symposium on Synthetic Media Attribution and Detection (ISSMAD) 2026}
\headerlogos{figures/resemble-logo}

\author{%
\textbf{Nicolas M.\ M\"uller},\quad
  \textbf{Aditya Tirumala Bukkapatnam},\quad
  \textbf{Zohaib Ahmed}
  \\[4pt]
  {Resemble AI, Mountain View, CA, USA}\quad
  \\[2pt]
  \email{\{nicolas,aditya,zohaib\}@resemble.ai}
}

\date{}

\begin{document}
\maketitle

\begin{abstract}
We present Proteus, a framework developed at Resemble AI for automated
  robustness testing of our audio deepfake detection system. Given a detector,
  Proteus systematically searches over sequences of everyday audio
  transformations (codec transcoding, additive noise, reverberation,
  dynamic-range compression, and VoIP simulation) to find combinations that
  fool the detector while preserving speech quality. We propose two
  complementary search strategies: (1)~a breadth-first search that
  exhaustively maps augmentation effectiveness across the parameter space, and
  (2)~a Q-learning agent designed to efficiently discover deeper attack chains
  by exploiting structural patterns in the BFS data. We report findings from
  continuous deployment of Proteus against our production detector, showing
  that specific augmentation chains can reliably flip detection verdicts while
  preserving speech intelligibility and speaker identity. We discuss how these
  findings are used to harden the detector through targeted retraining.

\end{abstract}

\section{Introduction}
\label{sec:intro}

Audio deepfake detectors are increasingly deployed to protect voice authentication,
media verification, and call-center fraud prevention. Yet these systems are almost
exclusively evaluated on clean, in-distribution test sets that do not reflect the
signal-processing transformations audio undergoes in production: lossy codecs, VoIP
channels, background noise, dynamic-range compression, and combinations thereof.

Recent work has begun to quantify this gap.
Li et al.~\cite{li2025measuring} show that codec-based corruptions cause the largest
performance drops across multiple detector architectures.
Wu et al.~\cite{wu2024clad} demonstrate that even simple manipulations (volume
changes, fading, resampling) can bypass detectors without affecting human perception.
Shi et al.~\cite{shi2025benchmarking} report 12--18\% EER increases under realistic
communication conditions.
ASVspoof~5~\cite{asvspoof5} was the first challenge to incorporate adversarial
conditions at scale.
However, these evaluations are \emph{passive}: they measure degradation under a fixed
set of conditions. None systematically \emph{search} for the augmentation chains that
maximally exploit a detector's weaknesses.

In prior work~\cite{deepen}, we showed that simple signal-processing attacks (time
stretching, echo addition, noise injection) reliably fool both production detectors
and academic models. DeePen established the threat, but the process was manual: a
human analyst selects attacks, observes results, and iterates. This does not scale to
the combinatorial space of multi-step augmentation chains with varying parameters.

We present \textbf{Proteus}, a framework developed at Resemble AI that automates this
search. Proteus composes augmentation chains from a library of 35 augmentations
spanning 11 categories (codecs, noise, reverb, VoIP, filtering, dynamics, distortion,
environmental effects, music overlay, silence injection, and temporal manipulation),
yielding approximately 110 distinct augmentation variants when accounting for
hyperparameter settings. It evaluates them against a target detector via black-box API
queries; chains that degrade speech intelligibility or speaker identity beyond
configurable thresholds are pruned.
Since the combinatorial complexity is immense even for shallow chains
($110^3 \approx 1.3 \times 10^6$ candidates at depth~3 alone), efficient search is
essential. We present two complementary search strategies: (1)~a breadth-first search
(BFS) that exhaustively maps single- and multi-step augmentation effectiveness, and
(2)~a Q-learning agent, described in Section~\ref{sec:qlearner}, designed to
scale the search to deeper chains by learning which augmentation transitions
are most effective.

In this paper, we describe the framework, both search strategies, and report findings from
continuous deployment against Resemble AI's production
detector,\footnote{\url{https://www.resemble.ai/detect/}} identifying specific
multi-step chains that evade detection while preserving perceptual quality.

\section{Method}
\label{sec:method}

Proteus operates as a black-box testing framework: it applies augmentation
chains to audio samples, queries a target detector via API, and records the
resulting detection scores. The framework consists of three components: (1)~a
composable augmentation library, (2)~a quality gate that enforces perceptual
constraints, and (3)~search strategies that navigate the augmentation space.

\subsection{Augmentation Library and Quality Gate}
\label{sec:auglib}

The augmentation library comprises 35 augmentation types organized into 11
categories: codecs (MP3, AAC, Opus, GSM, AMR), additive noise (white, pink,
babble), reverberation, filtering (low-pass, high-pass, band-pass), dynamics
(compression, limiting), distortion, temporal manipulation (time-stretch,
pitch-shift), environmental effects, music overlay, silence injection, and VoIP
simulation. Each augmentation type defines a set of discrete hyperparameter
configurations (\emph{variants}); for example, MP3 transcoding offers variants
at bitrates 32, 64, 96, 128, and 192\,kbps. The full library yields
approximately 110 distinct variants.

An augmentation chain $c = (a_1, a_2, \ldots, a_d)$ of depth~$d$ transforms
input audio $x$ into $\hat{x} = a_d \circ \cdots \circ a_1(x)$.  Because
arbitrary compositions can produce unintelligible or unrecognizable audio, every
chain must pass a \emph{quality gate} before the detector is queried.  The gate
enforces two constraints: word error rate (WER) between the original and augmented
transcripts (via Whisper~\cite{whisper}) must remain below a threshold
$\tau_{\text{wer}}$, ensuring intelligibility is preserved; and speaker
similarity (cosine similarity between speaker embeddings) must
exceed $\tau_{\text{spk}}$, ensuring the target speaker remains recognizable.
Together, these constraints scope the search to augmentations that constitute
realistic attacks: an adversary may want to disguise a deepfake as genuine
audio (impersonation, misinformation) or discredit authentic evidence as
synthetic, the so-called \emph{liar's dividend}~\cite{chesney2019deepfakes}.
In either direction, the attack is only useful if the content and speaker
identity survive the manipulation. Chains that fail the gate are discarded
without querying the detector.

\subsection{Breadth-First Search}
\label{sec:bfs}

The BFS strategy performs a level-wise exhaustive search. At level~1, every
augmentation variant is applied independently to each sample. Results are ranked
by absolute score shift $|\Delta s|$ from the unaugmented baseline, and the
top-$K$ chains are retained. At level~$\ell > 1$, each surviving chain from
level~$\ell{-}1$ is extended with every compatible variant (excluding immediate
self-repetition), yielding up to $K \times V$ candidates per sample, where~$V$
is the total number of variants. Early stopping terminates the search when score
improvements plateau across consecutive levels.

BFS provides a comprehensive map of augmentation effectiveness that reveals
which individual augmentations and pairwise transitions cause the largest
detector degradation. However, its complexity grows as $O(V^d)$ with chain
depth~$d$: at depth~3 with 110 variants, BFS must evaluate ${\sim}1.3 \times
10^6$ candidates per sample.

\subsection{Scaling to Deeper Chains: Q-Learning Agent}
\label{sec:qlearner}

To efficiently explore deeper chains, we formulate augmentation sequencing as a
Markov decision process (MDP)~\cite{puterman1994markov}. The key insight is that an augmentation's
effectiveness often depends on the \emph{preceding} augmentation: for instance,
adding noise \emph{before} a low-bitrate codec forces the encoder to spend
bits on the noise, compounding distortion of the speech signal, whereas
reversing the order simply layers noise on already-compressed audio. We exploit this sequential structure with a Q-learning
agent~\cite{sutton2018rl}.

\paragraph{MDP formulation.}
States correspond to augmentation \emph{types} (abstracting over
hyperparameters): $\mathcal{S} = \{s_0\} \cup \{s_1, \ldots, s_K\}$, where
$s_0$ is the initial (unaugmented) state and each $s_k$ represents the last
applied augmentation type ($K{=}35$). The action space $\mathcal{A} = \{a_1,
\ldots, a_K\}$ selects the next augmentation type. This yields a Q-table of
size $36 \times 35 = 1{,}260$ entries, small enough to learn from limited
evaluation data.

Each step involves two levels of decision: (1)~the Q-learner selects an
augmentation \emph{type} via upper confidence bound (UCB) action selection:
\begin{equation}
  a^* = \arg\max_{a} \left[
    Q(s, a) + c \sqrt{\frac{\ln N(s)}{N(s, a)}}
  \right],
  \label{eq:ucb}
\end{equation}
where $N(s)$ and $N(s, a)$ are visit counts and $c$ controls exploration; and
(2)~a per-augmentation categorical bandit samples a specific hyperparameter
variant, with probabilities updated based on quality outcomes and detector
feedback.

The reward $r$ is the marginal detector score shift from adding the current
step.  We use undiscounted ($\gamma{=}1$) Q-learning updates, since the
quality gate already limits chain depth:
\begin{equation}
  Q(s, a) \leftarrow Q(s, a) + \alpha
  \bigl[ r + \max_{a'} Q(s', a') - Q(s, a) \bigr].
  \label{eq:qupdate}
\end{equation}
On quality failure, a small negative reward is applied via the standard
update rule~\eqref{eq:qupdate}, gradually depressing Q-values for
augmentations that consistently fail the gate while preserving recovery
if the type later succeeds at less aggressive settings. The variant
bandit additionally decreases the chosen setting's probability.

\paragraph{Warmstart from BFS.}
The Q-table is initialized from available BFS results: for each observed
$(s, a)$ transition, the Q-value is set to the mean score shift.  Unseen
state-action pairs receive an optimistic default, ensuring the agent
preferentially explores untested transitions. This lets the Q-learner
exploit BFS findings without merely replaying them.

We present this formulation as a natural extension motivated by the BFS
results in Section~\ref{sec:results}: the strong dependence of augmentation
effectiveness on preceding steps suggests that an RL agent can learn
transition structure that BFS enumerates but cannot generalize from.
Experimental evaluation of the Q-learning agent is ongoing.

\section{Results}
\label{sec:results}

We report findings from a BFS run of Proteus against Resemble AI's production
deepfake detector.
It outputs a score in $[0,1]$
where values near~0 indicate bonafide and near~1 indicate spoof.
The search evaluates augmentation chains of depth~2 and~3
over eight baseline audio samples (four bonafide utterances from
M-AILABS~\cite{mailabs} and four spoofed utterances from
MLAAD~\cite{mlaad}), yielding 17{,}405 candidate chains in total. The quality
gate ($\tau_{\text{wer}} = 0.15$, $\tau_{\text{spk}} = 0.80$) rejects
12{,}558 candidates (72\,\%), leaving 4{,}847 chains that preserve
intelligibility and speaker identity.
Table~\ref{tab:top-chains} presents selected chains.
Depth-3 chains reveal interaction effects invisible to single-step
evaluations: in the second row, automatic gain control contributes only $+0.13$
at step~2, yet conditions the signal such that MP3 at step~3 produces its
largest observed marginal shift ($+0.51$).

\paragraph{False-positive vulnerability.}
The most striking finding is a strong asymmetry: \emph{all} of the top-100
chains (ranked by absolute score shift) target bonafide samples. Genuine
speech is far easier to push \emph{toward} the spoof boundary than synthetic
speech is to push \emph{away} from it. The best depth-2 chain,
\texttt{synthetic\_reverb} $\to$ \texttt{opus\_codec}, shifts a bonafide score
by $+0.99$, flipping the verdict. This confirms that the detector's primary
attack surface is false-positive induction, directly enabling the
\emph{liar's dividend}~\cite{chesney2019deepfakes}: an adversary can discredit
authentic recordings with modest signal processing that makes them appear
synthetic. We conjecture a distributional overlap: our augmentations' acoustic
by-products may be perceptually similar to the glitches of early-generation TTS,
causing the detector to mistake degradation for synthesis.
However, why certain combinations are disproportionately effective remains an open question. 

\begin{table}[t]
  \centering
  \caption{Selected high-shift augmentation chains on bonafide audio, chosen
    for diversity. Each row is one chain: Steps~1--3 list the augmentations
    applied in sequence (hyperparameter values withheld and denoted $\theta$), $d$~is chain depth, and
    $\Delta_i$~the marginal detector score shift at step~$i$;
    $\Sigma\!\Delta$~is the cumulative shift. For example, the first row
    applies room simulation, shifting the score by
    $+0.39$; MP3 transcoding adds $+0.33$; spectral gating then adds
    $+0.24$, yielding a cumulative shift of $+0.96$, nearly flipping the
    bonafide verdict to spoof. All chains pass quality gates
    (WER\,$< 0.15$, speaker similarity\,$> 0.80$).}
  \label{tab:top-chains}
  \small
  \begin{tabular}{@{}lllc |rrrr@{}}
    \toprule
    Step 1 & Step 2 & Step 3 & $d$ & $\Sigma\!\Delta$ & $\Delta_1$ & $\Delta_2$ & $\Delta_3$ \\
    \midrule
    room\_sim\,($\theta$) & mp3\,($\theta$) & spec\_gate\,($\theta$) & 3 & 0.96 & 0.39 & 0.33 & 0.24 \\
    room\_sim\,($\theta$) & auto gain\,($\theta$) & mp3\,($\theta$) & 3 & 0.79 & 0.16 & 0.13 & 0.51 \\
    \midrule
    synth\_reverb\,($\theta$) & opus\,($\theta$) & --- & 2 & 0.99 & 0.78 & 0.21 & --- \\
    static\_bg\,($\theta$) & pink\_noise\,($\theta$) & --- & 2 & 0.99 & 0.71 & 0.28 & --- \\
    echo\,($\theta$) & pkt\_loss\,($\theta$) & --- & 2 & 0.93 & 0.77 & 0.15 & --- \\
    music\_real\,($\theta$) & synth\_reverb\,($\theta$) & --- & 2 & 0.90 & 0.64 & 0.26 & --- \\
    \bottomrule
  \end{tabular}
\end{table}

\paragraph{Model hardening.}
The chains identified by Proteus feed directly into detector retraining.
High-shift chains are added to the training augmentation pipeline: bonafide
and spoofed samples are transformed with the discovered chains and included
as additional training data, teaching the detector to remain invariant to
these perturbations. After retraining, Proteus is re-run against the updated
model to verify that the targeted vulnerabilities are resolved and to surface
any new weaknesses. This creates a continuous adversarial testing loop in
which each Proteus run both validates prior fixes and drives the next round
of hardening.

\section{Conclusion}
\label{sec:conclusion}

We presented Proteus, a framework for automated adversarial robustness testing
of audio deepfake detectors. By searching over quality-constrained augmentation
chains, Proteus systematically identifies signal-processing conditions that
degrade detector performance.  BFS mapping provides a comprehensive robustness
profile, and the proposed Q-learning extension aims to scale this search to
deeper chains by learning augmentation transition structure.
Our deployment experience shows that adversarial robustness testing can be
embedded as a continuous practice in the model development lifecycle: findings
from Proteus directly inform augmentation strategies for detector retraining,
creating a closed loop between attack discovery and defense hardening.

\bibliography{references}

\begin{thebibliography}{11}
\providecommand{\natexlab}[1]{#1}
\providecommand{\url}[1]{\texttt{#1}}
\expandafter\ifx\csname urlstyle\endcsname\relax
  \providecommand{\doi}[1]{doi: #1}\else
  \providecommand{\doi}{doi: \begingroup \urlstyle{rm}\Url}\fi

\bibitem[Li et~al.(2025)Li, Chen, and Wei]{li2025measuring}
Xiang Li, Pin-Yu Chen, and Wenqi Wei.
\newblock Measuring the robustness of audio deepfake detectors.
\newblock \emph{arXiv preprint arXiv:2503.17577}, 2025.

\bibitem[Wu et~al.(2024)Wu, Chen, Du, Wu, He, Shang, Ren, and Xu]{wu2024clad}
Haolin Wu, Jing Chen, Ruiying Du, Cong Wu, Kun He, Xingcan Shang, Hao Ren, and
  Guowen Xu.
\newblock {CLAD}: Robust audio deepfake detection against manipulation attacks
  with contrastive learning.
\newblock \emph{arXiv preprint arXiv:2404.15854}, 2024.

\bibitem[Shi et~al.(2025)Shi, Shi, Dogan, Alzubi, Huang, and
  Zhang]{shi2025benchmarking}
Haohan Shi, Xiyu Shi, Safak Dogan, Saif Alzubi, Tianjin Huang, and Yunxiao
  Zhang.
\newblock Benchmarking audio deepfake detection robustness in real-world
  communication scenarios.
\newblock \emph{arXiv preprint arXiv:2504.12423}, 2025.

\bibitem[Wang et~al.(2024)Wang, Delgado, Tak, Jung, Shim, Todisco, Kukanov,
  Liu, Sahidullah, Kinnunen, Evans, Lee, and Yamagishi]{asvspoof5}
Xin Wang, H{\'e}ctor Delgado, Hemlata Tak, Jee-weon Jung, Hye-jin Shim,
  Massimiliano Todisco, Ivan Kukanov, Xuechen Liu, Md~Sahidullah, Tomi~H.
  Kinnunen, Nicholas Evans, Kong~Aik Lee, and Junichi Yamagishi.
\newblock {ASVspoof} 5: Crowdsourced speech data, deepfakes, and adversarial
  attacks at scale.
\newblock In \emph{Proc. ASVspoof Workshop}, 2024.

\bibitem[M{\"u}ller et~al.(2025)M{\"u}ller, Kawa, Stan, Doan, Jung, Choong,
  Sperl, and B{\"o}ttinger]{deepen}
Nicolas~M. M{\"u}ller, Piotr Kawa, Adriana Stan, Thien-Phuc Doan, Souhwan Jung,
  Wei~Herng Choong, Philip Sperl, and Konstantin B{\"o}ttinger.
\newblock {DeePen}: Penetration testing for audio deepfake detection.
\newblock \emph{arXiv preprint arXiv:2502.20427}, 2025.

\bibitem[Radford et~al.(2023)Radford, Kim, Xu, Brockman, McLeavey, and
  Sutskever]{whisper}
Alec Radford, Jong~Wook Kim, Tao Xu, Greg Brockman, Christine McLeavey, and
  Ilya Sutskever.
\newblock Robust speech recognition via large-scale weak supervision.
\newblock In \emph{Proceedings of the 40th International Conference on Machine
  Learning}, 2023.

\bibitem[Chesney and Citron(2019)]{chesney2019deepfakes}
Robert Chesney and Danielle~Keats Citron.
\newblock Deep fakes: A looming challenge for privacy, democracy, and national
  security.
\newblock \emph{California Law Review}, 107:\penalty0 1753--1819, 2019.

\bibitem[Puterman(1994)]{puterman1994markov}
Martin~L. Puterman.
\newblock \emph{Markov Decision Processes: Discrete Stochastic Dynamic
  Programming}.
\newblock Wiley, 1994.

\bibitem[Sutton and Barto(2018)]{sutton2018rl}
Richard~S. Sutton and Andrew~G. Barto.
\newblock \emph{Reinforcement Learning: An Introduction}.
\newblock MIT Press, 2nd edition, 2018.

\bibitem[Solak(2019)]{mailabs}
Imdat Solak.
\newblock The {M-AILABS} speech dataset.
\newblock 2019.
\newblock URL
  \url{https://www.caito.de/2019/01/03/the-m-ailabs-speech-dataset/}.

\bibitem[M{\"u}ller et~al.(2024)M{\"u}ller, Kawa, Choong, Casanova, G{\"o}lge,
  M{\"u}ller, Syga, Sperl, and B{\"o}ttinger]{mlaad}
Nicolas~M. M{\"u}ller, Piotr Kawa, Wei~Herng Choong, Edresson Casanova, Eren
  G{\"o}lge, Thorsten M{\"u}ller, Piotr Syga, Philip Sperl, and Konstantin
  B{\"o}ttinger.
\newblock {MLAAD}: The multi-language audio anti-spoofing dataset.
\newblock In \emph{International Joint Conference on Neural Networks (IJCNN)},
  2024.

\end{thebibliography}

\end{document}